\documentstyle[prl,aps,multicol,epsfig]{revtex}
\begin{document} 
\draft 
\title{Hubbard Fermi surface in the doped paramagnetic insulator}
\author{C. Gr\"ober, M. G. Zacher and R. Eder}
\address{Institut f\"ur Theoretische Physik, Universit\"at W\"urzburg,
Am Hubland,  97074 W\"urzburg, Germany}
\date{\today}
\maketitle

\begin{abstract}
We study the electronic structure of the doped paramagnetic insulator by 
finite temperature Quantum Monte-Carlo simulations for the 2D Hubbard model. 
Throughout we use the moderately high temperature $T=0.33t$, where the spin 
correlation length has dropped to $< 1.5$ lattice spacings, and study the 
evolution of the band structure with hole doping. The effect of 
doping can be best described as a rigid shift of the chemical potential into 
the lower Hubbard band, accompanied by a transfer of spectral weight. For hole
dopings $<20$\% the Luttinger theorem is violated, and the Fermi surface 
volume, inferred from the Fermi level crossings of the `quasiparticle band',
shows a similar doping dependence as predicted by the Hubbard I and related 
approximations. 
\end{abstract}
\begin{figure}
\epsfxsize=6.0cm
\vspace{-0.5cm}
\hspace{-0.5cm}\epsfig{file=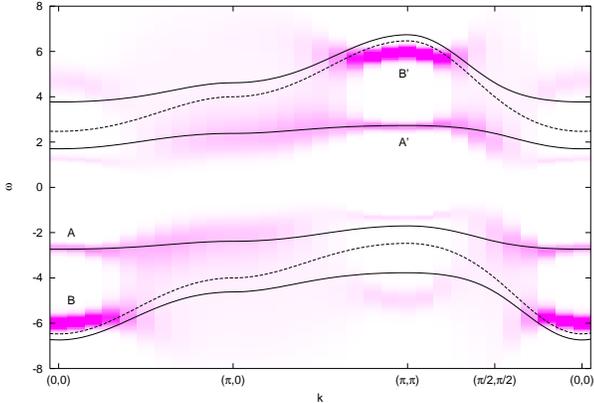,height=8.0cm,angle=270.0}
\vspace{-0.cm}
\narrowtext
\caption[]{Gray-scale plot
of the single particle spectral density for the $20\times 20$
cluster at half-filling. The gray scale gives the intensity
of spectral weight at the respective $\bbox{k}-\omega$ point.
Also shown are the Hubbard I bands (dashed lines)
and the $4$-bands obtained by solving (\ref{hybby}) (full lines).}
\label{fig1} 
\end{figure}

\begin{figure}
\epsfxsize=6.0cm
\vspace{-0.50cm}
\hspace{-0.0cm}\epsfig{file=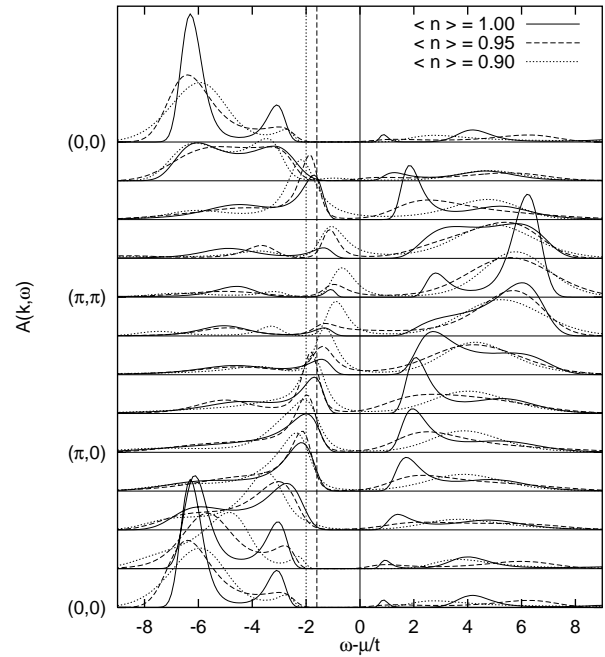,width=9.0cm,angle=270.0}
\vspace{0.cm}
\narrowtext
\caption[]{Single particle spectral function for the $8\times 8$
cluster and different electron density $\langle n\rangle$. 
The chemical potential
at half filling is the zero of energy, the spectra for different
$\langle n\rangle$ are rigidly shifted
relative to one another so as to match dominant features
The chemical potentials for the different $\langle n\rangle$
are given by vertical lines.}
\label{fig2} 
\end{figure}

 
\pacs{71.30.+h,71.10.Fd,71.10.Hf} 
\begin{multicols}{2}
Since the pioneering works of Hubbard\cite{hubbard}, the metal-insulator
transition in a paramagnetic metal has been the subject of intense
study. Despite this, our theoretical understanding of this
phenomenon is quite limited. Hubbard's original solutions to the problem,
the so-called Hubbard I-III approximations, have recently faced some
criticism. One fact which is frequently held against 
his approximations or the closely related
two-pole approximations\cite{roth,nolting,bernhard,beenen}
is the difficulty to reconcile them with the
Luttinger theorem. This can hardly be a surprise
in that all of these approaches rely on splitting
the electron creation operator into two `particles' which are
exact eigenstates of the interaction term $H_U$:
\begin{eqnarray}
c_{i,\sigma}^\dagger &=&
c_{i,\sigma}^\dagger n_{i,\bar{\sigma}} + 
c_{i,\sigma}^\dagger (1-n_{i,\bar{\sigma}})
\nonumber \\
&=& \hat{d}_{i,\sigma}^\dagger + \hat{c}_{i,\sigma}^\dagger,
\end{eqnarray}
with
$ [H_U, \hat{d}_{i,\sigma}^\dagger\;] = U \hat{d}_{i,\sigma}^\dagger$,
and
$[ H_U, \hat{c}_{i,\sigma}^\dagger\;] = 0$.
The interaction term is therefore treated exactly, 
approximations are made to the kinetic energy. This is
precisely the opposite situation as compared to the perturbation
expansion in $U$, which leads to the Luttinger theorem.
At half-filling the two `particles' $\hat{d}_{i,\sigma}^\dagger$ and
$\hat{c}_{i,\sigma}^\dagger$, whose energy of formation
differs by $U$,  then form the two separate Hubbard bands.
The effect of doping in both the Hubbard I approximation or
the two-pole approximations consists in the chemical potential cutting
gradually into the top of the lower Hubbard band, in much the same
fashion as in a doped band insulator. On the other hand the spectral weight
along the lower Hubbard band deviates from the
free-particle value of $1$ per momentum and spin so that the Fermi surface
volume (obtained from the requirement that the integrated
spectral weight up to the Fermi energy be equal to the total number 
of electrons) is not in any `simple' relationship to the
number of electrons - the Luttinger theorem must be violated.\\
In this manuscript we wish to address the question as to what really
happens if a paramagnetic insulator is doped away from half-filling, by
a Quantum Monte Carlo (QMC) study of the 2D Hubbard model. 
We use the value $U/t=8$ and
work throughout at the moderately high temperature $T=0.33t$.
This temperature is small compared to both the bandwidth, $U$,
and the gap in the single particle spectrum (see Figure \ref{fig1}).
The main effect of $T$ is
the destruction of antiferromagnetic order, as discussed
in our previous paper\cite{carsten1}.
We therefore believe that our study realizes to good approximation
the situation for which Hubbard's solutions
were originally designed: a paramagnetic system in the limit
of large $U$, at a temperature which is small
on the relevant energy scales.\\
Below, we present results for the single particle spectral function
and its doping dependence. These data show, that the
Hubbard I approximation is in fact considerably better than
commonly believed: the effect of doping indeed consists mainly of
a progressive shift of the chemical potential $\mu$ into
the band structure of the insulator.
The Fermi surface volume, if
determined in an `operational' way from the single particle spectral
function, indeed is not consistent with the Luttinger theorem.\\
We start with a brief
discussion of the band structure at half-filling, see Figure \ref{fig1},
which  shows the single particle spectral 
function. We note that this is quite consistent with previous 
QMC work\cite{Preuss}.
For comparison the two bands predicted by
the Hubbard I approximation,
\begin{equation}
E_{\pm}(\bbox{k}) = \frac{1}{2}\;[\;
(\epsilon_{\bbox{k}} + U) \pm 
\sqrt{\epsilon_{\bbox{k}}^2 + U^2}\;] 
\label{hubby}
\end{equation}
are also shown as the dashed dispersive lines
($\epsilon_{\bbox{k}}$$=$$-2t(\cos(k_x)+\cos(k_y)\;)$ is the
noninteracting dispersion).
These provide at best a rough fit to those parts
of the spectral function which have high spectral weight.
Inspection of the numerical spectra shows
quite a substantial difference between the numerical and the Hubbard-type
band structures: the latter always give two bands, whereas in the
numerical spectra one can rather unambiguously
identify $4$ of them, denoted as
as $B$, $A$, $A'$ and $B'$ (see the Figure). None of these
bands shows any indication of antiferromagnetic symmetry;
together with the short spin correlation length\cite{carsten1} 
\noindent 
this shows that we are really in the paramagnetic phase.
We found that to model this $4$-band structure one can
inroduce two additional dispersionless bands
at energies of $\bar{E}_{\pm}=\frac{U}{2}\pm \epsilon$.
We now allow mixing between each of these 
dispersionless bands and the respective Hubbard band, as would be described 
by the Hamilton matrix
\begin{eqnarray}
H_{\pm} =\left(
\begin{array}{c c}
E_{\pm}(\bbox{k}) & V \\
V & \bar{E_{\pm}}(\bbox{k})
\end{array} \right).
\label{hybby}
\end{eqnarray}
Using the values  $\epsilon=3t$ and $V=t$,
the resulting $4$-band structure provides at least a qualitatively
 correct fit to the numerical data.
We stress that at present we have no `theory' for these two
additional bands. Equation (\ref{hybby}) is just a
phenomenological {\em ansatz}
to fit the numerical band structure. We note, however,
that a $4$-band structure which has some similarity
with our results has recently been 
obtained by Pairault {\em et al.} using a strong 
coupling expansion\cite{pairault}.\\
We do not, however, pursue this issue
further but turn to our main subject, the effect of hole doping.
Figure \ref{fig2} shows the development of $A(\bbox{k},\omega)$
with doping. Thereby the  $A(\bbox{k},\omega)$ for different hole
concentrations have been overlaid so as to match dominant features,
and the chemical potentials for the different hole 
concentrations are marked by lines.
It is quite obvious from this Figure that the $2$ bands seen at half-filling
in the photoemission spectrum persist with an essentially unchanged
dispersion. The chemical potential gradually cuts deeper and deeper
into the $A$ band, forming a hole-like Fermi surface centered
on $(\pi,\pi)$, the top of the lower Hubbard band.
The only deviation from a rather simple rigid-band behavior is
an additional transfer of spectral weight:
the part of the $A$-band near $(\pi,\pi)$ gains in spectral weight,
whereas the $B$-band looses weight. The loss of the
$B$ band cannot make up for the increase of the
$A$ band, but rather there is an additional transfer of weight
from the upper Hubbard bands, predominantly the
$A'$ band. This effect is quite well understood\cite{henk}.
The $A'$ band seems to be affected strongest by
the hole doping and in fact the rather clear
two-band structure visible near $(\pi,\pi)$ at half-filling
rapidly gives way to one broad `hump' of weight.
Apart from the spectral weight transfer, however, the
band structure on the photoemission side is almost unaffected
by the hole doping - the {\em dispersion} of the
$A$-band becomes somewhat wider but does not change appreciably. 
In that sense we see at least qualitatively the behavior
predicted by the Hubbard I approximation.
\noindent 
Next, we focus on the Fermi surface volume. Some care
is necessary here: first,
we cannot actually be sure that at the high temperature we are using there
is still a well-defined Fermi surface. Second, the criterion we will be using
is the crossing of the $A$ band
through the chemical potential. It has to be
kept in mind that this may be quite misleading, because band portions with
tiny spectral weight are ignored in this approach (see for example
Ref. \cite{comment} for a discussion).
When thinking of a Fermi surface as the constant energy contour of the
chemical potential, we have to keep in mind that
portions with low spectral weight may be overlooked.
On the other 
hand the fact that a peak with appreciable weight crosses 
from photoemission to inverse photoemission
at a certain momentum is independent of whether we call this a
`Fermi surface' in the usual sense, and should be reproduced by
\begin{figure}
\epsfxsize=6.0cm
\vspace{-1.0cm}
\hspace{1.0cm}\epsfig{file=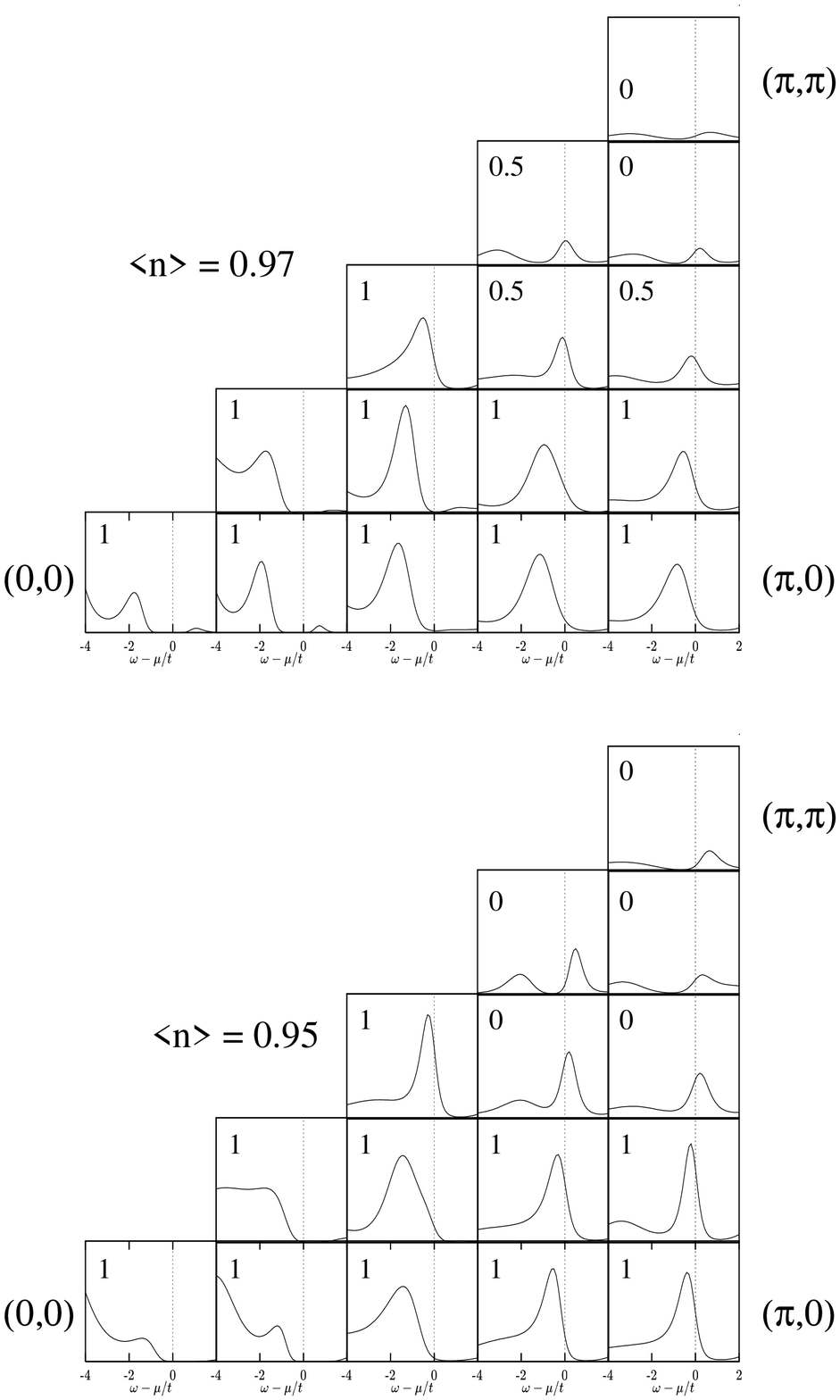,width=7.4cm,angle=0.0}
\vspace{0.0cm}
\narrowtext
\caption[]{Single particle spectral function for all
${\bf k}$-points of the $8\times 8$ cluster in the irreducible
wedge of the Brillouin zone. For each ${\bf k}$ the weight
$w_{\bf k}$ is given.}
\label{fig3} 
\end{figure}
\begin{figure}
\epsfxsize=6.0cm
\vspace{-1.5cm}
\hspace{1.0cm}\epsfig{file=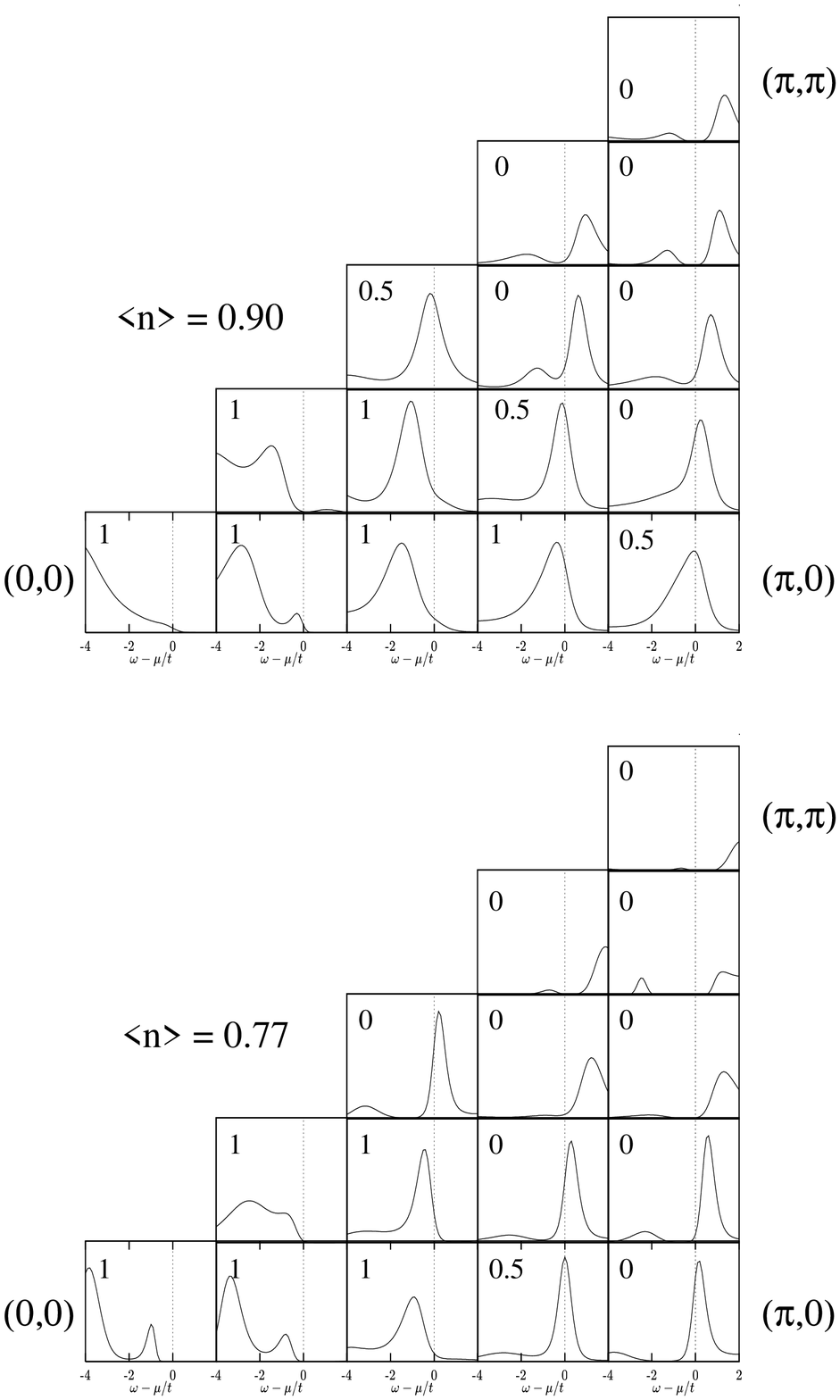,width=7.4cm,angle=0.0}
\vspace{0.0cm}
\narrowtext
\caption[]{Same as Figure \ref{fig3} for lower electron densities.}
\label{fig4} 
\end{figure}
\noindent 
any theory which claims to describe the system.
It therefore has to be kept in mind that in the following
we are basically studying a `spectral weight Fermi surface', i.e.
the locus in ${\bf k}$ space where an apparent quasiparticle band
with high spectral weight crosses the chemical potential.
With these {\em caveats } in mind, Figures \ref{fig3} and
\ref{fig4} show the low-energy
peak structure of $A(\bbox{k},\omega)$ for all allowed momenta
of the $8\times 8$ cluster in the
irreducible wedge of the Brillouin zone, and 
for different hole concentrations.
In all of these spectra there is a pronounced peak,
whose position shows a smooth dispersion with momentum.
Around $(\pi,\pi)$ the peak is clearly above $\mu$, whereas in the
center of the Brillouin zone it is below.
The locus in $\bbox{k}$-space where the peak crosses $\mu$ forms
a closed curve around $(\pi,\pi)$ and it is obvious
from the Figure that the `hole pocket' around
$(\pi,\pi)$ increases very rapidly with
$\delta$. To estimate the Fermi surface volume $V_F$
we assign a weight $w_{\bf k}$ of $1$ to momenta ${\bf k}$ where the
peak is below $\mu$, $0.5$ if the peak is right at $\mu$ and
$0$ if the peak is above $\mu$. Our assignments of these weights
are given in Figure \ref{fig3}. 
The fractional Fermi surface volume then is
$V_F = \frac{1}{N}\sum_{\bf k} w_{\bf k}$, where $N=64$ is the
number of momenta in the $8\times 8$ cluster.
Of course, the assignment of the $w_{\bf k}$
involves a certain degree of arbitrariness.
It can be seen from Figures
\ref{fig3} and \ref{fig4},however, that our  $w_{\bf k}$ 
\begin{figure}
\epsfxsize=6.0cm
\vspace{-0.5cm}
\hspace{1.5cm}\epsfig{file=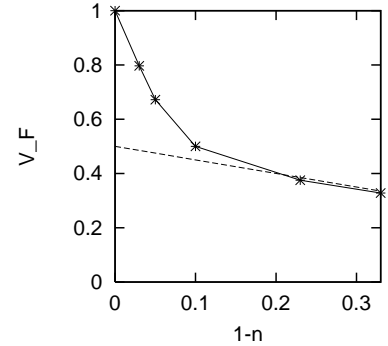,width=5.0cm,angle=270.0}
\vspace{-0.0cm}
\narrowtext
\caption[]{Fermi surface volume as estimated from the
single particle spectral function, plotted versus
the concentration of holes in the half-filled band.
The dashed line gives the value predicted by the Luttinger theorem,
$V_F=\frac{n}{2}$.}
\label{fig5} 
\end{figure}
\noindent 
would in any way tend to underestimate the Fermi surface volume, so 
that the obtained $V_F$ data points rather have the character of a 
lower bound to the true $V_F$. Even if we take into account some
small variations of $V_F$ due to different assignments
of the weight factors, however, the resulting $V_F$ versus $\delta$ 
curve never can be made consistent with the Luttinger volume, see
Figure \ref{fig5}. The deviation from the Luttinger 
volume is quite pronounced at low doping.
$V_F$ approaches the Luttinger volume for dopings
$\approx 20$\%, but due to our somewhat crude way of
determining $V_F$ we cannot really decide when precisely 
the Luttinger theorem is obeyed. The Hubbard I approximation
approaches the Luttinger volume for hole concentrations of $\approx 50$\%,
i.e. the steepness of the drop of $V_F$ is not reproduced
quantitatively. The latter is somewhat improved in 
the so-called $2$-pole approximation\cite{bernhard,beenen}. For
example the Fermi surface given by
Beenen and Edwards\cite{beenen} for $\langle n \rangle=0.94$ obviously
is very consistent with the spectrum in Figure \ref{fig3} for
$ \langle n \rangle=0.95$.\\
In summary, we have studied the doping evolution of the single particle
spectral function for the paramagnetic phase
of the 2D Hubbard model, starting out from the insulator.
As a surprising result, we found that in this situation the
Hubbard I and related approximations give a qualitatively
quite correct picture. The main discrepancy between the
Hubbard I and the so-called $2$-pole
approximation and our numerical spectra is the number
of `bands' of high spectral weight, which is
$4$ in the numerical data. This is no reason for concern,
because we have seen that adding two more bands
allows for an quite reasonable fit to the
numerical band structure and one might expect that finding a
somewhat more intricate decoupling scheme for the Hubbard I approximation
or a suitable $4$-pole approximation should not pose a major problem.
The greatest success of the Hubbard-type approximations, however,
is a qualitatively quite correct description
of the evolution of the `Fermi surface'.
The effect of doping consists of the progressive shift of the
chemical potential into the topmost band observed at half-filling,
accompanied by some transfer of spectral weight.
The Fermi surface volume, determined in an `operational way' from the
band crossings, violates the Luttinger theorem for
low hole concentrations and does not
appear to be in any simple relationship
to the electron density. The Luttinger sum rule is recovered only
for hole concentrations around $20$\%. \\
It is interesting to note in this context that
a recent study of the momentum distribution in the
t-J model by high-temperature series expansion\cite{Puttika}
has also provided some evidence for a `Fermi surface' which encloses
a larger value than predicted by the Luttinger theorem.
The criterion used there was a maximum of $|\nabla_{\bbox{k}} n_{\bbox{k}}|$,
i.e. the locus of the steepest drop of $ n_{\bbox{k}}$. This would in fact 
be quite consistent with the present results. However, the same
{\em caveat} as in the present case applies, i.e. this
criterion will overlook Fermi level crossings of bands with low
spectral weight\cite{comment}.\\
In our opinion the strange dependence of $V_f$ on electron density
makes it questionable whether the `spectral weight Fermi surface'
in our data is a true constant energy contour for
a system of `quasiparticles'. It may be possible that at the
temperature we are studying a Fermi surface in the usual
sense no longer exists, and that
the Hubbard I approximation merely reproduces
the {\em spectral weight distribution} in this case.
As our data show, however, for that purpose the approximation is considerably
better than commonly believed.\\
Zero temperature studies for the doped
t-J and Hubbard model are only possible
by using exact diagonalization\cite{dagoreview}, 
in which case the shell-effects
due to the small system size
require special care\cite{shimozato,nishimoto}.
One crucial point is the very different shape of the
quasiparticle dispersion at zero temperature.
Whereas the $A$ band is at least topologically equivalent to
a {\em nearest neighbor} hopping dispersion, with minimum at
$(0,0)$ and maxiumum  at $(\pi,\pi)$, the zero temperature 
data\cite{dagoreview} show a {\em second-nearest neighbor}
dispersion with a nearly degenerate band maximum along the
antiferromagnetic zone boundary, and a shallow
absolute maximum at $(\pi/2,\pi/2)$.
The effect of hole doping at zero temperature, however,
has a qualitatively very similar effect as in the present 
case\cite{shimozato}:
the chemical potential simply cuts into the quasiparticle band
for the insulator, which thus is populated by hole-like 
quasiparticles\cite{nishimoto}.
Again, these `hole pockets' violate the Luttinger theorem, indicating
again the breakdown of adiabatic continuity in the low doping regime
persists also at low temperatures.\\
We thank W. Hanke for useful comments.
This work was supported by DFN Contract No. TK 598-VA/D03, by BMBF
(05SB8WWA1),
computations were performed at HLRS Stuttgart, LRZ M\"uchen and HLRZ J\"ulich.

\end{multicols}

\begin{references}
\bibitem{hubbard} 
J. Hubbard, Proc. Roy. Soc. A {\bf 276}, 238 (1963);
J. Hubbard, Proc. Roy. Soc. A {\bf 277}, 237 (1964);
J. Hubbard, Proc. Roy. Soc. A {\bf 281}, 401 (1964).
\bibitem{roth}
 L. M. Roth,  Phys. Rev. {\bf 184}, 451 (1969).
\bibitem{nolting}
 G. Geipel and  W. Nolting, Phys. Rev. B {\bf 38}, 2608 (1988);
  W. Nolting and W. Borgiel, Phys. Rev. B {\bf 39}, 6962 (1989).
\bibitem{bernhard}
 B. Mehlig, H. Eskes, R. Hayn, and M. B. J. Meinders,
 Phys. Rev. B {\bf 52}, 2463 (1995).
\bibitem{beenen}
 J. Beenen and D. M. Edwards, Phys. Rev. B {\bf 52}, 13636 (1995).
\bibitem{carsten1}
 C. Gr\"ober, M. G. Zacher, and R. Eder, cond-mat/9810246.
\bibitem{Preuss}
 R. Preuss, W. Hanke and W. von der Linden,
    Phys. Rev. Lett. {\bf 75}, 1344 (1995).
\bibitem{pairault}
 S. Pairault, D. Senechal, and A.-M. S. Tremblay,
 Phys. Rev. Lett. {\bf 80}, 5389 (1998).
\bibitem{henk}
 H. Eskes and A. M. Oles, Phys. Rev. Lett. {\bf 73} 1279 (1994).
\bibitem{comment}
 R. Eder and Y. Ohta, Phys. Rev. Lett. {\bf 72}, 2816 (1994).
\bibitem{Puttika}
 W. O. Putikka, M. U. Luchini,
 and R. R. P. Singh, Phys. Rev. Lett.  {\bf 81}, 2966 (1998).
\bibitem{dagoreview}
 E. Dagotto, Rev. Mod. Phys. {\bf 66}, 763 (1994).
\bibitem{shimozato}
 R. Eder, Y. Ohta, and T. Shimozato, Phys. Rev. B {\bf 50}, 3350 (1994);
 R. Eder and Y. Ohta, Phys. Rev. B {\bf 51}, 6041 (1994).
\bibitem{nishimoto}
 S. Nishimoto, Y. Ohta, and R. Eder, Phs. Rev. B {\bf 57}, R5590 (1998).
\end{references}
\end{document}